\documentclass{XrU2005}
\usepackage{graphicx}
\usepackage{natbib}

\title{The Extended Chandra Deep Field-South Survey: Optical Properties of X-ray Detected Sources}
\author[1]{Shanil Virani}
\author[1,2]{Ezequiel Treister}
\author[1]{C. Megan Urry}
\author[1]{Eric Gawiser}
\author[3]{the MUSYC Collaboration}
\affil{Department of Astronomy, Yale University}
\affil[2]{Departamento de Astronomia, Universidad de Chile}
\affil[3]{http://www.astro.yale.edu/musyc}

\newcommand{\chandra}{\mbox{\em Chandra\/}}

\begin{document}


\maketitle

\begin{abstract}
\vspace{-0.8cm}
The Extended \chandra\ Deep Field-South (E-CDF-S) survey
consists of 4 \chandra\ ACIS-I pointings and covers $\approx$1100 square 
arcminutes ($\approx$0.3 deg$^2$) centered on the original CDF-S field
to a depth
of approximately 228 ks (PI: Niel Brandt; \citealp{lehmer}). 
This is the largest \chandra\ survey ever conducted 
at such depth. In our analysis \citep{virani},
we detect 651 unique sources --- 587 using a conservative source detection 
limit and 64 using a lower source detection limit. Of these 651 sources, 561 
are detected in
the full 0.5--8.0 keV band, 529 in the soft 0.5--2.0 keV band, and 335
in the hard 2.0--8.0 keV band. 
For point sources near the
aim point, the limiting fluxes are approximately $1.7 \times
10^{-16}$ $\rm{erg~cm^{-2}~s^{-1}}$ and $3.9 \times
10^{-16}$ $\rm{erg~cm^{-2}~s^{-1}}$ in the 0.5--2.0 keV and
2.0--8.0 keV bands, respectively. We present the optical properties of these
X-ray sources, specifically the $R$-band magnitude distribution and a preliminary spectroscopic redshift distribution. One exciting result is the discovery of
 7 new Extreme X-ray-to-Optical 
flux ratio objects (EXOs) found in the E-CDF-S field.

\end{abstract}

\vspace{-0.8cm}
\section{Source Detection}
\vspace{-0.8cm}

\begin{figure}[t]
\includegraphics[width=0.42\textwidth, angle=0]{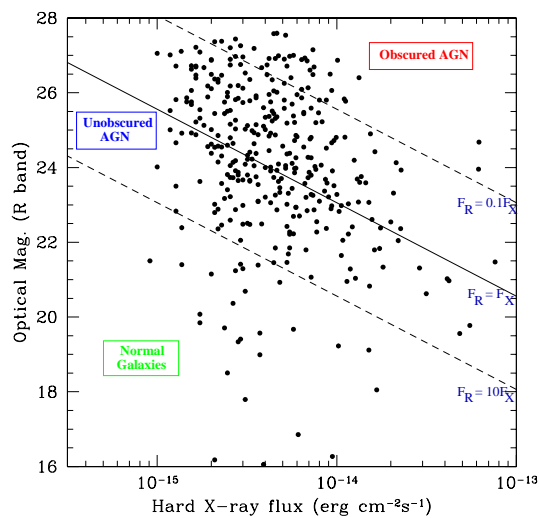}
\caption{2--8 keV flux vs. $R$-band magnitude (AB) for sources in the E-CDF-S.
Unobscured AGN typically populate the region between the dashed lines, while
obscured AGN typically lie above this region, and ``normal'' galaxies lie below
this region. \label{hfvsr}}
\end{figure}

We report on the sources detected in three
standard X-ray bands: 0.5--8.0 keV (full band), 0.5--2.0 keV (soft band),
and 2.0--8.0 keV (hard band). To perform X-ray source detection, we applied 
the CIAO
wavelet detection algorithm \textit{wavdetect} using a 
``$\sqrt{2}$~sequence'' of wavelet scales; scales of 1, $\sqrt{2}$, 2, 
$2\sqrt{2}$, 4, $4\sqrt{2}$, and 8 pixels were used. Our criterion for 
source detection is that a source must be found with a false-positive 
probability threshold ($p_{thresh}$) of $1\times 10^{-7}$ in at least one 
of the three standard bands. We also produced a second catalog using a more
liberal probability threshold of $1\times 10^{-6}$. This scheme resulted in
a total of 651 unique X-ray sources detected in the E-CDF-S survey field \citep{virani}\footnote{X-ray catalog and images available at http://www.astro.yale.edu/svirani/ecdfs}. 
Figure~\ref{hfvsr} presents the hard X-ray flux versus the $R$-band magnitude.

\begin{table} [ht]   
\caption{\textbf{MUSYC-ECDFS 5$\sigma$~ AB Point Source Limits}\label{limits}}
\begin{center}       
\begin{tabular}{cccccccc} 
\hline
\hline

\rule[-1ex]{0pt}{3.5ex} BVR & U & B & V & R & I & z & NB5000 \\
\hline\hline

27.1 & 26.0 & 26.9 & 26.4 & 26.4 & 24.6 & 23.6 & 25.5 \\
\hline\hline
\end{tabular}
\end{center}
\end{table}

\vspace{-0.8cm}
\section{Multiwavelength Survey by Yale-Chile (MUSYC)}
\vspace{-0.8cm}

MUSYC is a square-degree survey to AB limiting depths of U,B,V,R=26 and K=22
(K=23 in the central 10\arcmin$\times$10\arcmin ~of each field), with extensive follow-up spectroscopy \citep{gawiser}. Table~\ref{limits} lists the 5$\sigma$
point source limits in each of the filters for the E-CDF-S field.
The project comprises four 30\arcmin$\times$30\arcmin ~fields, of
which the E-CDF-S is one. Ground-based imaging has been completed and deep follow-up spectroscopy (to R$\sim$25) is underway (Magellan/IMACS, VLT/VIMOS, Gemini/GNIRS). 

\begin{figure}[!t]
\includegraphics[width=0.42\textwidth]{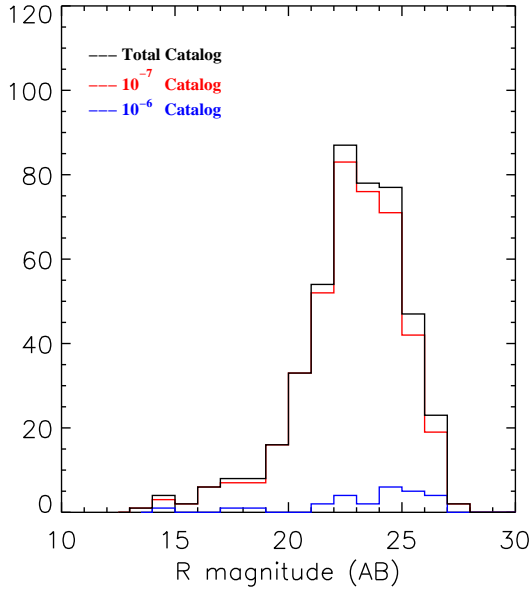}
\caption{$R$-band magnitude distribution for the optical counterparts to the X-ray detected sources. Most of the sources have $R$-band magnitudes~ $<$ 25 mag making them suitable for spectroscopic follow-up. \label{Rbandhist}}
\end{figure}

\begin{figure}[!t]
\includegraphics[width=0.42\textwidth]{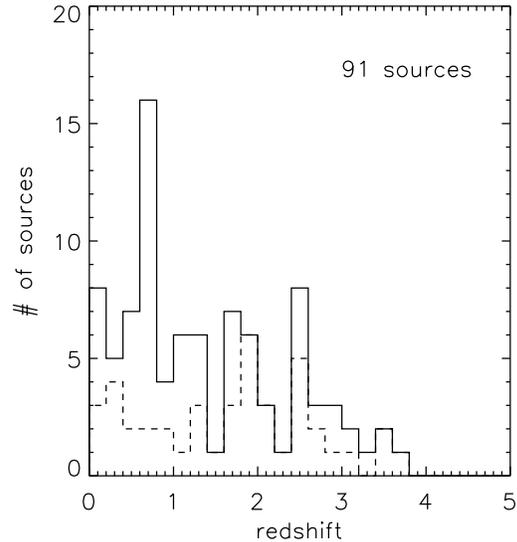}
\caption{Spectroscopic redshifts for 91 X-ray sources in the E-CDF-S (solid line). 
This plot includes $\sim$50 redshifts from the \citet{szokoly} catalog, 
as well as $\sim$40 redshifts determined from Magellan/IMACS spectroscopy (dashed line) performed as part of the MUSYC survey. \label{zhist}}
\end{figure}

\vspace{-0.8cm}
\section{Optical Counterparts and Spectroscopy}
\vspace{-0.8cm}

\begin{figure}[t]
\includegraphics[width=0.42\textwidth, angle=0]{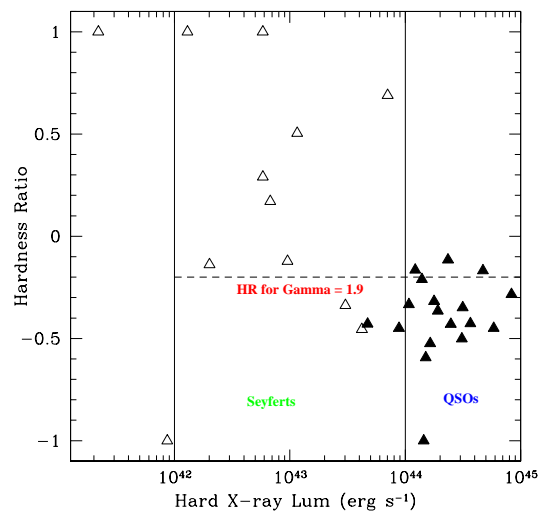}
\caption{Hardness ratio vs. the hard X-ray luminosity for the $\sim$40 sources for which spectroscopic redshifts were determined as part of the MUSYC survey. Early results indicate that luminous AGN are less obscured; more obscured AGN at low luminosities. This is partly a selection effect and partly real.
\label{hrvslum}}
\end{figure}

In the primary (10$^{-7}$) catalog,  420 out of 587 sources (72\%) have a unique optical counterpart within 1.5$''$ of the X-ray source position (3 X-ray sources have multiple optical counterparts) in the deep MUSYC catalog of 84,410 $BVR$-selected sources.  
In the secondary (10$^{-6}$) catalog, 26 out of 64 sources (41\%) have unique optical counterparts. Figure~\ref{Rbandhist} shows the $R$-band magnitude distribution for these sources. Figure~\ref{zhist} shows the spectroscopic redshift distribution for 91 X-ray sources obtained thus far. Most are broad line, unobscured AGN (Figure~\ref{hrvslum}). 
Additional spectroscopic observation runs are scheduled. 
Lastly, there are 7 sources, referred to as Extreme X-ray-to-Optical flux ratio objects (EXOs; \citealp{koekemoer}), that are undetected in the deep $BVR$ imaging but are robustly detected in MUSYC $K$-band imaging of this field (Virani et al, in prep).

%
\end{document}